\documentclass[twocolumn]{aastex61}
\usepackage{url}
\usepackage{graphicx}

\newcommand\etal{~et~al.}

\begin{document}
\title{Compact Resolved Ejecta in the Nearest Tidal Disruption Event}

\author{Eric S. Perlman}
\affiliation{Department of Physics and Space Sciences, Florida Institute of Technology, 150 W. University Blvd., Melbourne, FL 32901, USA}

\author{Eileen T. Meyer}
\affiliation{Department of Physics, University of Maryland -- Baltimore County, 1000 Hilltop Circle, Baltimore, MD  21250, USA}

\author{Q. Daniel Wang}
\affiliation{Department of Astronomy, University of Massachusetts, LGRT-B 619E, 710 North Pleasant Street, Amherst, MA 01003-9305, USA} 

\author{Qiang Yuan}
\affiliation{Department of Astronomy, University of Massachusetts, LGRT-B 619E, 710 North Pleasant Street, Amherst, MA 01003-9305, USA}

\affiliation{Purple Mountain Observatory, Chinese Academy of Sciences, no. 2 West Beijing Road, Nanjing, Jiansu 210008, China}

\author{Richard Henriksen}
\affiliation{Department of Physics, Engineering Physics \& Astronomy, Queens University, Kingston, Ontario, K7L 3N6, Canada}

\author{Judith Irwin}
\affiliation{Department of Physics, Engineering Physics \& Astronomy, Queens University, Kingston, Ontario, K7L 3N6, Canada}

\author{Marita Krause}
\affiliation{Max-Planck Institut f\"ur Radioastronomie, Auf dem H\"ugel 69, D-53121, Bonn, Germany} 

\author{Theresa Wiegert}
\affiliation{Department of Physics, Engineering Physics \& Astronomy, Queens University, Kingston, Ontario, K7L 3N6, Canada}

\author{Eric J. Murphy}
\affiliation{US Planck Data Center, The California Institute of Technology, MC 220-6, Pasadena, CA 91125, USA }

\author{George Heald}
\affiliation{CSIRO Astronomy and Space Science, 26 Dick Perry Avenue, Kensington WA 6151, Australia}
\affiliation{Kapteyn Astronomical Institute, PO Box 800, 9700 AV, Groningen, The Netherlands}

\author{Ralf-J\"urgen Dettmar}
\affiliation{Astronomical Institute, Faculty for Physics and Astronomy, Ruhr-Universit\"at Bochum, D-44780 Bochum, Germany}
\begin{abstract}

Tidal disruption events (TDEs) occur when a star or sub-stellar object passes close enough 
to a galaxy's supermassive black hole to be disrupted by tidal forces. NGC 4845 (d=17 Mpc) 
was host to a TDE, IGR J12580+0134, detected in November 2010. Its proximity offers us a unique close-up of the TDE and its 
aftermath.  We discuss new Very Long Baseline Array
({\sl VLBA}) and Karl G. Jansky Very Large Array ({\sl JVLA}) {\footnote{The National Radio Astronomy Observatory 
and Long Baseline Observatory are facilities of the 
National Science Foundation operated under cooperative agreement by Associated Universities, Inc.}} observations,  
which show
that the radio flux from the active nucleus created by the TDE has decayed in a manner consistent with
predictions from a jet-circumnuclear medium interaction model.   This model explains the source's broadband spectral 
evolution, which shows a spectral peak that has moved from the submm (at the end of 2010) to GHz radio frequencies 
(in 2011-2013) to $<1$ GHz in 2015. The milliarcsecond-scale core is 
circularly polarized at 1.5 GHz but not at 5 GHz, consistent 
with the model. The {\sl VLBA} images show a complex structure at 1.5 GHz that includes an east-west extension $\sim 40$ milliarcsec (3 pc) long as well as a resolved component 52 milliarcsec (4.1 pc) northwest of the flat-spectrum core, which 
is all that can be seen at 5 GHz.  If ejected in 2010, the NW component must have had  
$v=0.96 c$ over five years.  However, this is unlikely, as our model suggests  strong deceleration to speeds 
$< 0.5c$ within months and a much 
smaller, sub-parsec size.  In this interpretation, the northwest component could have either a non-nuclear origin or be from an earlier event.

\end{abstract}

\section{Introduction}

A star or sub-stellar object will be partially or fully tidally disrupted
when it passes close enough by a supermassive black hole (SMBH). Such
tidal disruption events (TDEs) are expected to occur every $10^3-10^5$
years for a typical galaxy (Magorrian \& Tremaine 1999; Wang \& Merritt
2004). This rate could be substantially higher (as high as once every 
few years), if the SMBH has a companion which is either an intermediate-mass black hole or SMBH (e.g., Chen et al. 2009). The debris of the disrupted
object will be accreted onto the black hole, producing flaring emission
at X-ray, ultraviolet, and optical wavelengths. A typical $t^{-5/3}$ behavior
of the X-ray luminosity, following the decrease of the fallback rate of the
debris, is a distinctive feature of TDEs (Phinney 1989). Jets can also be launched
by such an event. When they interact with the circum-nuclear medium (CNM)
high energy particle acceleration could occur (e.g., Cheng et al. 2006).
Observations show that at least some TDEs do launch relativistic jets.
Sw J1644+57 (full name Swift J164449.3+573451, $z=0.3534$; Bloom et al. 2011; Burrows et al. 2011; Levan et al. 2011; Zauderer et al. 2011)
and Sw J2058+05 ($z=1.1853$; Cenko et al. 2012) are examples, exhibiting
super-Eddington X-ray emission and a long lasting radio emission
expected to arise from the jet-CNM interaction. Detailed modeling of both
events suggests that the jets were moving along our lines of sight. It is
then natural to expect that there should be more events with off-axis jets.

IGR J12580+0134 was a TDE detected in the nucleus of NGC 4845 -- a galaxy located in the Virgo cluster, at a
distance of only $\sim17$ Mpc. Due to its proximity, IGR 1258+0134 gives a rare chance to  scrutinize a TDE and its
aftermath with the highest possible resolution.   The source was initially detected in  2010 November
by {\it Integral} (Walter et al. 2011). Follow-up X-ray observations with
{\it XMM-Newton}, {\it Swift} and {\it MAXI}, together with {\it
Integral} data, suggest that the source probably resulted from a TDE of
a super-Jupiter by the galaxy's central SMBH (Nikolajuk \& Walter 2013).
Spectral fitting to the {\it XMM-Newton} data indicates a soft X-ray
excess (with temperature 0.33 keV) above the power-law (with index $\Gamma
\simeq2.2$) from the TDE likely representing the collective emission from
unrelated discrete sources and/or truly diffuse hot plasma because of the
completely different absorption compared with that of the TDE (Nikolajuk 
\& Walter 2013). 

The radio counterpart of the TDE  was
detected serendipitously in 2011 December by the {\sl JVLA}
in a nearby galaxy survey (CHANG-ES; Irwin et al. 2015) where the core was a factor 10 brighter than seen in {\it FIRST} 
observations conducted between 1993-2004. The radio spectrum,
peaking at GHz frequencies, and its variation suggest self-absorbed
synchrotron emission with changing optical thickness. These phenomena
can be naturally explained by an expanding radio lobe, powered by the jet
injection of the TDE. A moderately relativistic jet model fits the original radio data (Irwin et al,2015) and predicts one interpretation of the current VLA and VLBA data. A relativistic jet model  gives a good fit to both the current and original data (Lei et al. 2016), but suggests a rather different interpretation of the VLBA data. 
From this model, we estimated the initial Lorentz factor
of the jet as $\Gamma_i\sim10$ and a viewing angle of 
$40^{\circ}$. This off-axis
viewing direction of the jet may
explain its sub-Eddington luminosity (Nikolajuk \& Walter 2013) as a result of a lower Doppler boosting factor.
Furthermore, an extended (diameter $\sim20$ arcsec) disk component with 
spectral index $\alpha=0.74 ~(F_\nu \propto \nu^{-\alpha})$ around the nucleus is shown in the radio data 
(Irwin et al. 2015), which is probably the counterpart of the soft X-ray 
excess and related to past star formation and/or SMBH activities. 

\section{Observations and Data Reduction}

\subsection{{\sl JVLA} Observations}

IGR J12580+0134 was observed five times with within a period of 7 months with the {\sl VLA} in 2011-2012 as part of the observations made of NGC 4845 during
the CHANG-ES program.  These observations, which allowed the detection of the radio emission from the TDE 
1-2 years from the event, were done at L band in the B, C and D 
configurations, and in C band in the C and D configurations.  At L-band, the frequency coverage was (in GHz) 
1.247 $\rightarrow$ 1.503 and 1.647 $\rightarrow$ 1.903 (500 MHz total); at C-band it was 4.979 $\rightarrow$ 7.021 (2 GHz). The frequency gap at L-band was set to avoid very strong persistent interference in that frequency range.  The data and basic reduction procedures were previously discussed in 
Irwin et al. (2015), but here we detail additional work that was done subsequent to the analysis in that
paper.  Details of those observations are given in Table 1.

We observed IGR J12580+0134 with the {\sl JVLA} in A-configuration at 
L-band (1.5 GHz) and C-band (5.5 GHz) on 22 June 2015 as part of program 15A-357. Both observations had a 
bandwidth of 1 GHz.  Total integration time 
(over a single scan) was 129 and 479 seconds, respectively. The observations were done in full polarization 
mode (8-bit setups L16f2A and C16f2A), with 26 antennas. At both bands, 3C 286 was used to calibrate the flux 
scale and bandpass and J1224+0330 was used to calibrate phase. For polarization calibration in both bands, 3C 
286 was used to set the angle. Unfortunately, an unpolarized calibrator was not observed multiple times on 22 June 2015 which would have been needed for a proper  correction of instrumental polarization.
For L-band, we used calibration scans on 3C 48 and 3C 84 from dataset 15A-305 (PI Werner) taken on 23 June 
2015. For this latter dataset, 3C 48 was used as a bandpass calibrator before generating the d-terms from 3C 
84. At C-band, the leakage terms were set by a single observation of 3C 84 from project 15A-252 (PI Murphy) 
taken on 15 August 2015.  The data reduction was conducted using CASA version 4.5.0 (release 35147). All 
datasets were Hanning smoothed and then inspected and flagged for RFI before a final calibration was applied 
following standard procedures. 

At L-band, after splitting off the source scan, we did a final flagging on the source scan and then imaged the data 
using CLEAN in multi-frequency synthesis (mfs) mode with nterms=2, with a wide field gridding mode and 1189 
w-projection planes (calculated by CLEAN), and briggs weighting (robust parameter 0.5). The image field was 
2700$\farcs$ square with 0.25$\farcs$ pixels. The image shown in Figure~1 (top) was generated using CLEAN in psfmode 
ÔCLARKÕ after two rounds of phase-only self-calibration.  However, to look for both linear and circular polarization, we also ran 
CLEAN on the original dataset with psfmode ÔCLARKSTOKESÕ and nterms=1.  No significant signal 
was found in Q, U, or V.  At C-band, we proceeded similarly to the L-band dataset, with an images size of 612\farcs 
square and 0.07\farcs pixels (beam size 0.47\farcs $\times $ 0.33\farcs), multi-frequency synthesis (mfs) mode with nterms=2, with a 
wide field gridding mode and 901 w-projection planes (calculated by CLEAN). The image shown in Figure~1 (bottom) is 
the I-only image after 5 rounds of phase-only self-calibration.  Again, we also conducted a full stokes imaging 
(with nterms=1) for the C-band dataset but found no significant signal in Q, U or V.

\begin{deluxetable*}{lclccccc}
\tablecolumns{8}
\tabletypesize{\scriptsize}

\tablewidth{0pt}
\tablecaption{Radio Interferometry Observations.}

\tablehead{ 
\multicolumn2c{Observation}&&& \multicolumn{4}{c}{RMS Noise, $\mu$Jy beam$^{-1}$} \\
\colhead{Date (Epoch)} & \colhead{Array \& Config.} & \colhead{$\nu$ (GHz)} & \colhead{Beam Size (PA), arcsec} & 
\colhead{$I$} & \colhead{$Q$} & \colhead{$U$} & \colhead{$V$} }
\startdata
1995 Feb 27 & VLA/D (NVSS) & 1.4 & $45 \times 45 (0)$ & 450 & ... & ... & ... \\
1997 Aug 21, 25 & VLA/C & 8.4 & $7.08 \times 5.69 (64.51)$ &  100 & ... & ... & ... \\
1998 Oct 09 & VLA/B (FIRST) & 1.4351 & $6.4 \times 5.4 (0)$ & 150 & ... & ... & ... \\
\hline
2011 Dec 19 (T1) & VLA/D & 5.99833 & $10.98 \times 9.06 (-1.40) $& 15 & 15 & 15 & 12 \\
2011 Dec 30 (T1) & VLA/D & 1.57470 & $38.58 \times 34.27 (-5.22)$ &40 & 27 & 27 & 28 \\
2012 Feb 23, 25 (T2) & VLA/C & 5.99854 & $3.05 \times 2.75 (-11.71)$& 3.9 & 3.2 & 3.2 & 3.3 \\
2012 Mar 30 (T2) & VLA/C  & 1.57484 & $12.18 \times 11.10 (-41.76)$ &45 & 19 & 19 & 26 \\
2012 Jun 11 (T3) & VLA/B & 1.57499 & $3.51 \times 3.33 (22.69)$ &18 & 15 & 15 & 15 \\
\hline
2015 Jun 22, 26 (T4) & VLA/A & 1.5195 & $1.73 \times 1.06 (43.27)$ & 88 & 58 & 58 & 59 \\ 
2015 Jun 22, 26 (T4) & VLA/A & 5.499 & $0.47 \times 0.33 (46.20)$ & 13 & 16 & 17 & 16 \\ 
2015 Oct 08 & VLBA & 1.5474 & $9.26^a \times 3.71^a (-2.83)$ &128 & 48 & 46 & 124 \\
2015 Oct 08 & VLBA & 4.9795 & $2.91^a \times 1.19^a (-0.52)$ & 55 & 37 & 40 & 35 \\
\enddata
\tablenotetext{a}{milliarcseconds}
\end{deluxetable*}

\subsection {{\sl VLBA} Observations}

We observed IGR J12580+0134 with the {]\sl VLBA} on 8 October 2015 (Table 1).
Observations were done at 1.5 GHz (L band) and 5.0 GHz (C band), 
using the Roach Digital Back End (RDBE) and 8 IFs.  Each IF is 32 MHz wide, and contains 128 $\times$ 250 kHz 
channels.  All 10 antennas of the {\sl VLBA} were used. 
The observations were done in full polarization mode, using 
 3C 286 to calibrate the {\sl JVLA}'s flux scale.  J1254+0233 
was used as a primary phase calibrator, and we used OQ208 as a D-term calibrator (all of these 
sources were observed on 8 October).  
Observations of NGC 4845 and J1254+0233 were grouped together in blocks of four (two at each frequency), with 
each individual scan on NGC 4845 lasting 4 minutes.  Observations of OQ 208 were done every 4th iteration,
allowing coverage of more than 80$^\circ$ in parallactic angle,
and the observations were phase referenced, allowing for maximum 
sensitivity as well as absolute positional referencing.  The total observing time was 5 hours, split evenly between
the two bands. The data were correlated at the {\sl VLBA} correlator in Socorro, New Mexico.

\begin{figure}[ht]
\includegraphics[width=8.5cm]{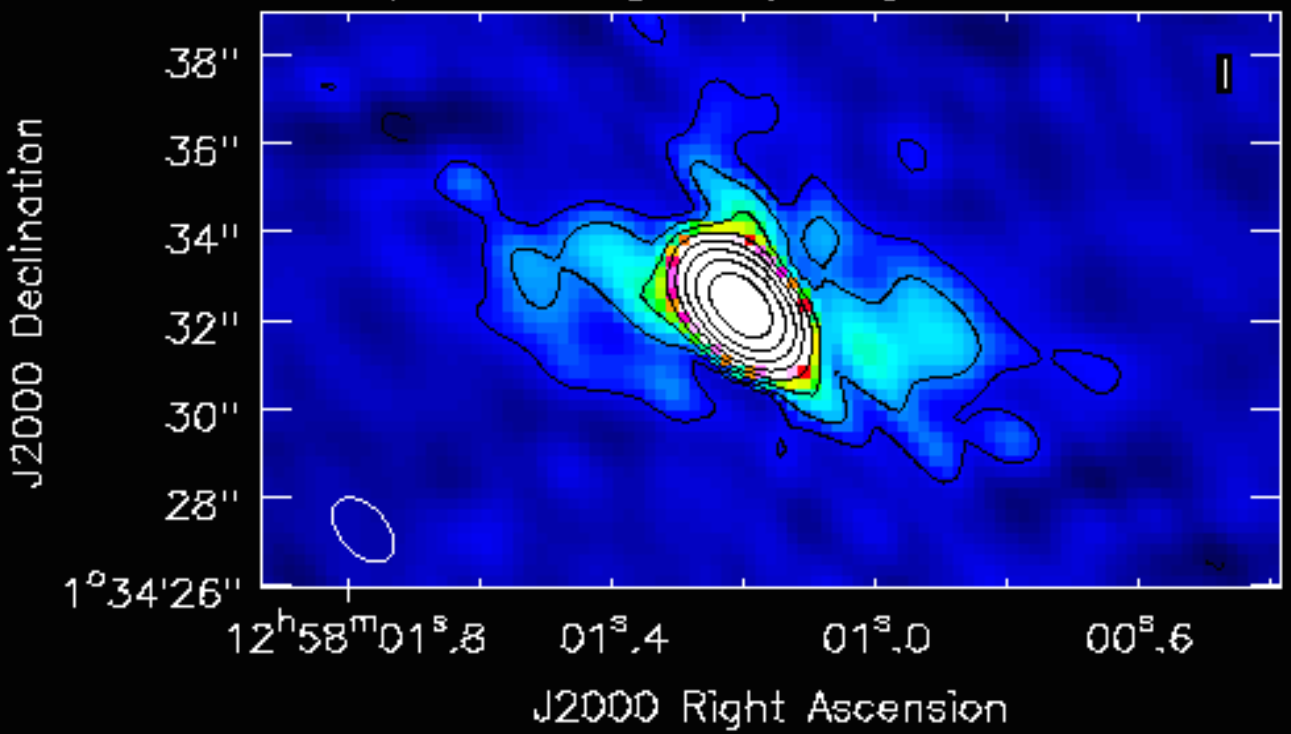}
\includegraphics[width=8.5cm]{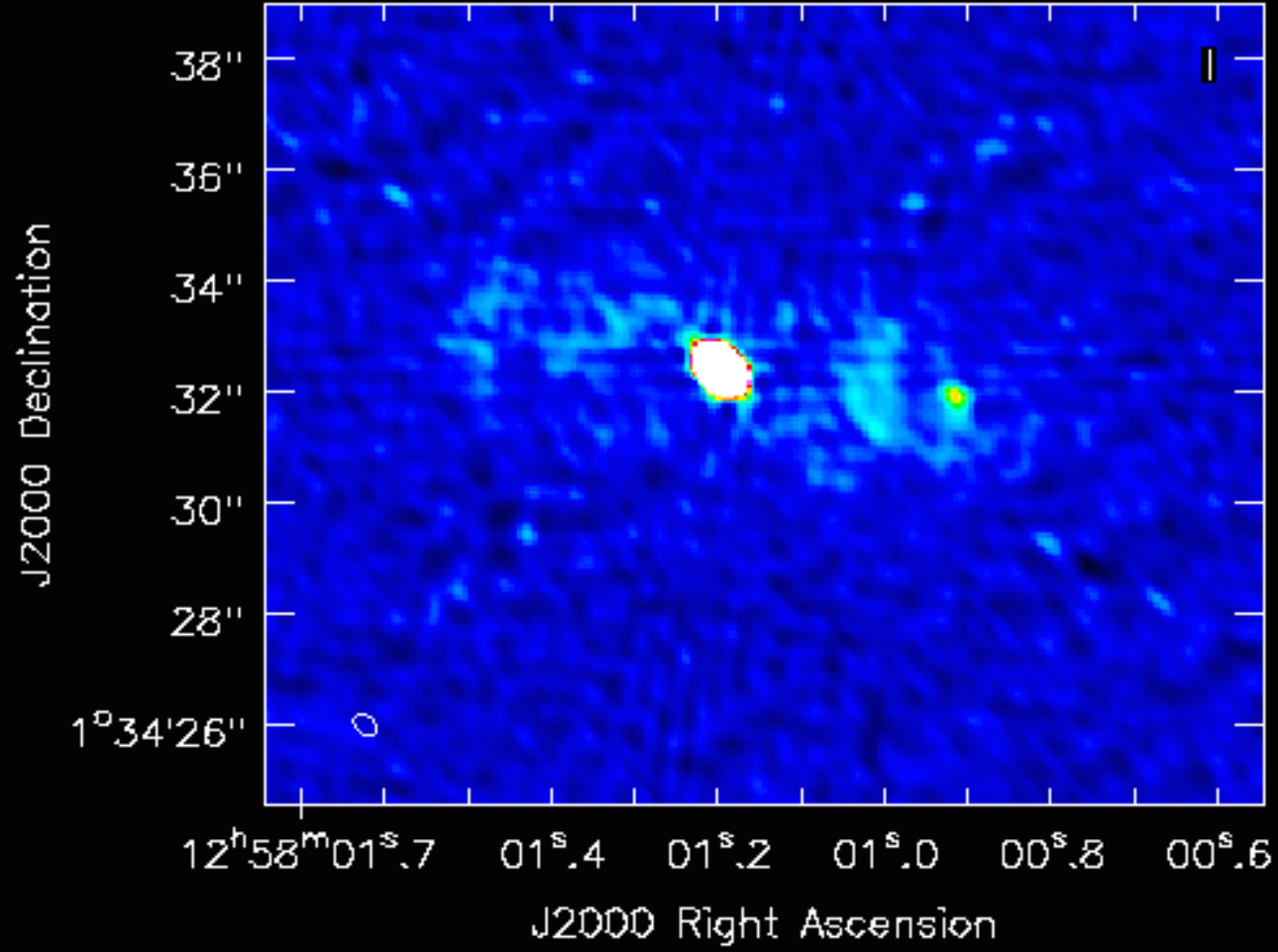}
\caption{Images of the nuclear region of NGC 4845 from the {\sl JVLA} in June 2015.  At top, we show the L band 
observation, while at bottom, we show the C band observation.  Note that both images show extended emission 
that is likely associated with the galaxy's disk, as  well as the nuclear component discussed in \S 3. The contours are at multiples of 4 times the off-source RMS.}
\end{figure}

All data reduction for the {\sl VLBA} observations were done in AIPS, following standard recipes in the AIPS 
Cookbook, specifically Appendix C{\footnote{http://www.aips.nrao.edu/cook.html}}.  
We  corrected for earth orientation, ionospheric conditions and sampler errors using VLBAEOPS, VLBATECR 
and VLBACCOR. Following this, we corrected for instrumental delays using VLBAPCOR and applied bandpass
calibration in VLBABPSS.  We then inspected the data in POSSM and did a priori amplitude calibration with 
VLBAAMP, using the gain and system temperature curves for each station, yielding correlated flux densities.  
Prior to fringe fitting, the supplied pulse calibration information was applied with the task PCCOR.  Following this,  
the data were fringe-fit in VLBAFRNG.  We also referenced the 
phases of NGC 4845 to those of J1254+0233.
~
This allowed 
us to find absolute positional information for all the sources, as well as improve the sensitivity of the 
observations (see e.g., Beasley \& Conway 1995, Wrobel 2000). Because of some fairly severe 
interference at L band that could not easily be flagged out of the data, 
the phase referencing was not successful in that band, and therefore 
we do not refer to absolute positions in that band.  Parallactic angle and 
polarization calibration was done using VLBAPANG  using the OQ208 and 3C 286 observations.  
We followed the recommendation of the NRAO staff on polarization calibration{\footnote{http://www.vla.nrao.edu/astro/
calib/polar/}}, using 3C 286 to calibrate the position angles.  Once these steps were done, the calibration
was applied in SPLIT, which also yields single-source data files.    To check the polarization calibration, we 
imaged J1254+0233 in both total flux and in polarization.  No circular polarization was found.

Hybrid-mapping procedures were started using a point-source model for initial phase calibration. 
All images used ROBUST=0 Briggs weighting (Briggs 1995).
In subsequent iterations of self-calibration, we allowed first the phase and then both amplitude and phase to 
vary.   We took care to ensure that each successive iteration of self-calibration did not go so deep as to start 
including negatives, residual side-lobes or other spurious values, and that the peak specific intensity did not 
decline.  In all, 4 iterations of phase-only self-calibration and 2 iterations of amplitude and phase self-calibration 
were done for the Stokes I images.   All imaging and self-calibration were performed in AIPS, using the
tasks IMAGR and CALIB, respectively, for the imaging/cleaning and self-calibration.
Imaging of Stokes Q, U and V were carried out using the self-calibration
tables developed in the Stokes I imaging, combined with the polarization and D-term calibrators. We 
did not find significant emission in either
Stokes $Q$ or $U$.  The circular polarization maps were made without fitting a spectrum across the band, 
because of the much smaller bandwidth 
of the {\sl VLBA} data as compared to the {\sl JVLA} data.  

\section{Results}

The radio flux densities detected in our {\sl JVLA} and {\sl VLBA} observations are given in Table 2, along with the flux 
densities and in-band spectral indices.  These include the 
values observed in 2011-2012 as part of the CHANG-ES survey (Irwin et al. 2015), as well as the 2015 {\sl JVLA}
imaging we report here.  To compare with the CHANG-ES data reported in Irwin et al. (2015) and plotted in 
Fig. 2, we define 22 June 2015 as ÒT4Ó, or T1 + 1270 days (times T1-T3 are defined in Irwin et al. 2015). 
 
\subsection{The AGN and its radio variability}

In the A-array observations we see evidence of a disk-like extension of diffuse emission, as 
previously seen in the earlier CHANG-ES data obtained with more compact configurations
although it is likely that some larger-scale 
structure is being resolved out at A-configuration.   We measured the flux of the central point 
source by fitting a gaussian profile. At L-band the resulting flux is $118 \pm 1$ mJy, with an in-band spectral
index of  $\alpha=0.54 \pm 0.02$. At C-band we 
measure $36.3 \pm 0.1$ mJy with an in-band spectral index of $\alpha=1.25 \pm 0.03$. The error on the in-band
spectral indices is  dominated by errors in the flux model of the {\sl JVLA}, as discussed in Perley \& Butler (2013, 2016).   The radio spectral 
index between the two bands is $\alpha=0.90$, indicating significant spectral curvature between the bands. 

\begin{deluxetable*}{lcccccccc}
\tabletypesize{\scriptsize}
\tablecolumns{9}
\tablewidth{0pt}
\tablecaption{Radio Interferometry Observations.}

\tablehead{
\colhead{Date} & \colhead{Config.} & \colhead{$\nu$ (GHz)}
&\colhead{$I_{tot}$, mJy} & \colhead{$I_{peak}$, mJy} & $\alpha$ &  \colhead{$P$, mJy} & \colhead{$V$, mJy} & \colhead{$V/I, \%$}
}
\startdata
1995 Feb 27 & VLA/D (NVSS) & 1.4 & $46.0 \pm 1$ \\
1997 Aug 21, 25 & VLA/C & 8.4 & 12.5  \\
1998 Oct 09 & VLA/B (FIRST) & 1.4351 & $33.9 \pm 0.4$  \\
\hline
2011 Dec 19 & VLA/D & 6.00 & $432 \pm 2$ & $425 \pm 3$ & & $2.3 \pm 0.9$ & $<0.036$ & $<8.3\times10^{-3}$ \\ 
2011 Dec 30 & VLA/D & 1.57 & $230 \pm 2$ & $211 \pm 3$ & & $0.55 \pm 0.05$ & $6.6 \pm 0.2$ & $2.9 \pm 0.1$\\
2012 Feb 23, 25 & VLA/C & 6.00 & $362 \pm 1$ & $355 \pm 1$ & $0.4 \pm 0.2$ &$ <0.01$ &$ <0.01$ & $<2.8\times10^{-3}$\\
2012 Mar 30 & VLA/C & 1.57 & $260 \pm 2$ & $241 \pm 3$ & & $0.35 \pm 0.03$ &$ 5.7 \pm 0.1$ & $2.2 \pm 0.1$  \\ 
2012 Jun 11 & VLA/B & 1.57 & $238 \pm 1$ & $219 \pm 4$ & & $1.2 \pm 0.5$ & $5.5 \pm 0.4$ & $2.3 \pm 0.2$ \\
\hline
2015 Jun 22, 26 & VLA/A & 1.5195 & $ 118.0 \pm 1.2$ & $114.5 \pm 0.7$ & $0.54 \pm 0.02$ & $<0.12$& $<0.12$ & $<0.1$\\
2015 Jun 22, 26 & VLA/A & 5.499 & $36.32 \pm 0.07$ & $36.00 \pm 0.04$ & $1.25 \pm 0.001$ & $<0.03$ & $<0.03$ & $<0.1$ \\
2015 Oct 08 & VLBA & 1.5474 & $83.5^a \pm 1.1 $ & $65.9 \pm 0.6$ & $ ...^b$& $<0.07$ & $1.4 \pm 0.2$  & $1.7 \pm 0.3$\\ 
2015 Oct 08 & VLBA & 4.9795 & $29.9 \pm 0.2 $ & $ 14.0 \pm 0.1$  & ...$^b$ & $<0.06$ & $<0.07$ & $<0.2$  \\
\enddata

\tablenotetext{a}{includes both core ($72.7 \pm 0.7$ mJy) and fainter (northern) source ($10.8 \pm 0.9$ mJy). }
\tablenotetext{b}{Band is too narrow for reliable fitting (\S 2.2).}

\end{deluxetable*}
\begin{figure}[hb]

\includegraphics[width=8.5cm]{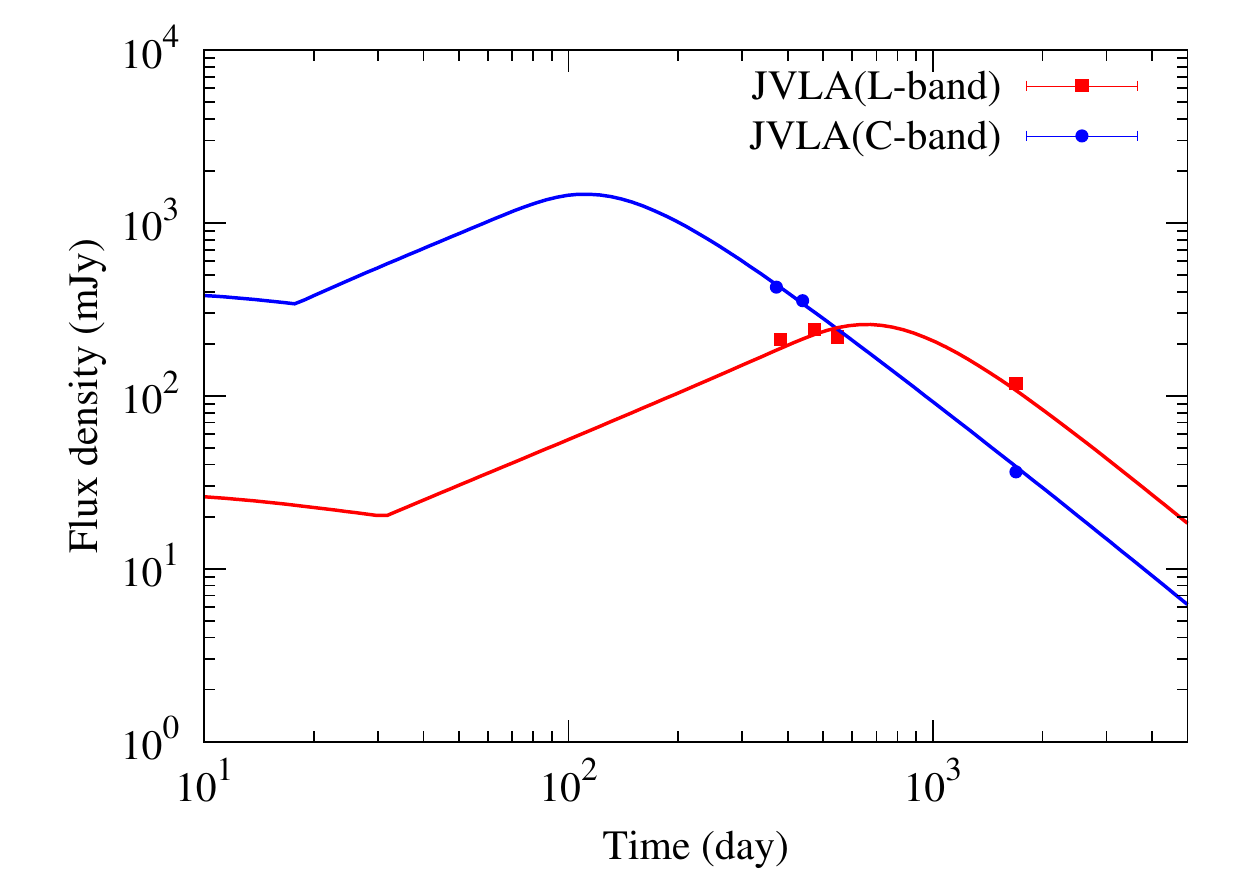}
\caption {Radio light curves of the TDE, compared to a jet-CNM interaction model (Lei et al.
2016, Irwin et al. 2015), which predicts a power-law ($t^{-5/3}$) decay.
Our data follow this very well, but note the importance of the 2015 data in obtaining the correct fit.}
\end{figure}

\begin{figure}[h]
\includegraphics[width=8.5cm]{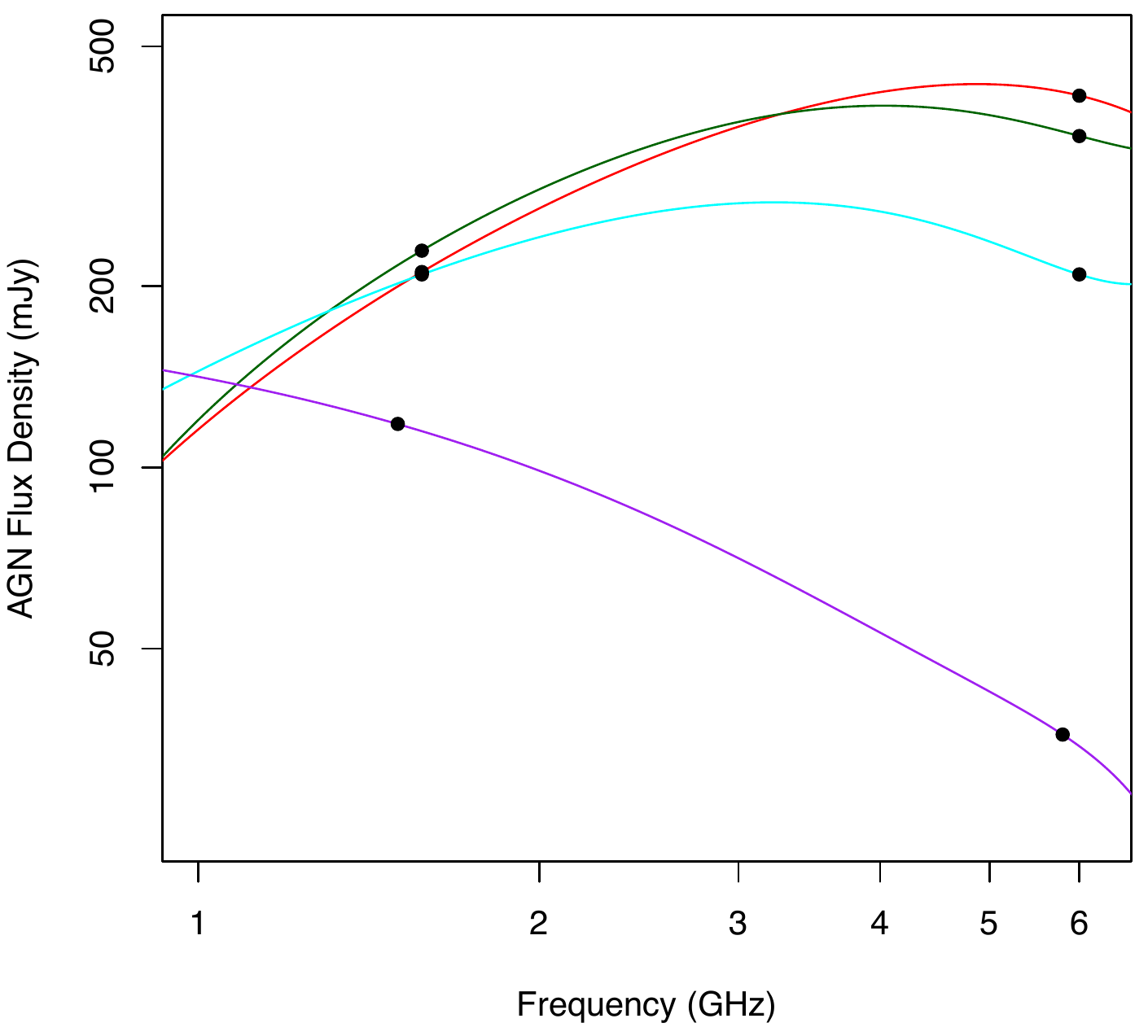}
\caption{Radio spectra of the nuclear source (Table 2), along with polynomial fits, for four time stamps:  $T_1$ 
(2011 Dec 30) in red, $T_2 = T_1+56$ days in green, $T_3 = T_1 + 196$ days in cyan, and $T_4 = T_1 + 1270$ 
days in purple.  Note that not only has the radio spectrum between 1 and 6 GHz changed from strongly inverted 
at epochs $T_1$ and $T_2$ to flat at $T_3$ and steep at $T_4$, but also the position of the spectral peak has 
evolved, from $\sim 5$ GHz at $T_1$ to $\sim 3.5$ GHz at $T_2$ to $< 1$ GHz at $T_4$.  Note that the 
apparent spectral upturn of the fit to the $T_4$ data is most likely an artifact and not real:  a power-law form 
(see bottom panel) is much more likely. } 
\end{figure}

Nearly five years from the initial event, it 
is clear that the radio source corresponding to the active nucleus (AGN) of NGC 4845
is still quite bright, albeit decreased significantly from that observed in late 2011, when flux densities 
$\sim 250$ mJy were observed. Multiple views of the light curve of the TDE are shown in Figure 2.  
Also looking at Table 2, it is obvious that the radio spectrum has evolved as a function of time, and the 
gigahertz peak seen in 2011-2012 has 
been replaced with a power law that steepens between 1 and 6 GHz.  
The radio spectrum has clearly evolved with time in the years since the 2010 November TDE.  
The 2015 {\sl JVLA} and {\sl VLBA} data are consistent with the presence 
of a spectral peak at frequencies below $\sim 1$ GHz, as predicted by the model, but we do not actually see 
any peak.  While this is most likely due to a lack of lower-frequency observations, we cannot exclude the 
alternate possibility of multiple spectral components.  We do not
include the  {\sl VLBA} observations in Figure 2, as their resolution differs by nearly 3 
orders of magnitude from that of the {\sl JVLA} observations so that the flux densities are not comparable.
This spectral evolution is shown in Figure 3. We discuss the evolution of the SED and its implications, in \S 4.2.

\subsection{Parsec-scale structure of the AGN}

The central source of NGC 4845 was detected at both L and C band with the VLBA.  We show the VLBA 
images in Figure 4.    The {\sl VLBA}
images show two sources at L band, separated by 51.7 milliarcsec, translating to 4.1 pc projected distance. 
 The main source is extended along the east-west direction, with the source being nearly 40 milliarcsec (3 pc) long 
and the flux maximum region being a broad plateau extending roughly SE to NW (Figure 4, bottom).
Only one source is seen at C band, however.  We fitted the flux maxima in both L and C band images with 
elliptical Gaussians using JMFIT.  The size and PA of these Gaussians were allowed to vary because of the 
extended nature of the central source in both images. The position of the C-band flux maximum 
is $\alpha={\rm 12h 58m 01.19814s,}$, $\delta = 01^\circ 34^\prime 32.4203^{\prime\prime}$,  
with internal $1\sigma$ errors (from JMFIT) of about 0.1 milliarcsec.  
The fitted component  at C band had a 
maximum intensity of $9.80 \pm 0.05$ mJy/beam and a integrated intensity of $30.2\pm 0.2$ mJy 
with a deconvolved full-width at half-maximum of $2.8 \pm 0.3 \times 2.3 \pm 0.2$ milliarcsec in PA $128^\circ$.
The fitted component at L band had a 
maximum intensity of $12.7 \pm 0.1$ mJy/beam and an integrated intensity of $60.1 \pm 0.6$ mJy 
with a deconvolved full-width at half-maximum of $14.8 \pm 0.2 \times 5.6 \pm 0.3$ milliarcsec in PA $103^\circ$.  
Since JMFIT fits a Gaussian, 
the  internal errors may be underestimates; however, they are consistent with the 
usual  expectations of a few tenths of a pixel (the pixel size was 1 milliarcsec for the L band 
{\sl VLBA} image and 0.3 milliarcsec for the C band {\sl VLBA} image).  
The total flux of the southern {\sl VLBA} source is $72.7 \pm 0.7$ mJy at 1.5 GHz
and $29.9 \pm 0.2$ mJy at 4.9 GHz, while the NW source, seen only at L band, has a flux of $10.8 
\pm 0.9$ mJy. The total flux detected in the {\sl VLBA} observations at L band is 71\% of that seen in the {\sl JVLA}
observation, while the total flux detected by the {\sl VLBA} at C band is 83\% of that seen in the {\sl JVLA} A-array data.
It is thus apparent that the majority of the flux seen by the {\sl JVLA} observations originated on milliarcsecond 
scales, although 20-30\% of that flux was resolved out and is on scales too large to be detected by the 
{\sl VLBA} observations.  Given the faintness of the source and the diffuse nature of the extended flux
it is difficult to trust the outer contours too heavily, particularly given the interference in L band.  
However, the overall east-west nature of the central source and the resolved, NW component 
survive all efforts to mask them out during cleaning.

While the spectral indices between L and C band are relatively easy to calculate in the {\sl JVLA} data, the resolved
nature of the {\sl VLBA} structure means one needs to take additional steps.  Using the entire {\sl VLBA} source
at both frequencies, without discriminating between the extended and unresolved components, one obtains 
$\alpha=0.95$, broadly consistent with the 2015 {\sl JVLA} inter-band spectral index.  However, 
if instead we calculate a spectral index for the southern VLBI source,
one obtains $\alpha=0.81$ if one uses the entire flux measurement at L band, or $\alpha=0.70$ using just the
unresolved L band flux.   This is probably more indicative of the core, and roughly consistent with what is seen on arcsecond scales at 
epoch $T_4$ (Table 2 and Figure 3).  While
formally the {\sl VLBA} data at L band cover over 300 MHz in bandwidth, in practice for this faint a source it proved impossible
to fit a good spectral index value over any significant part of the image (for example the spectral index derived for 
the central source is $0.67 \pm 0.59$).

\begin{figure}[ht]
\includegraphics[width=8.5cm]{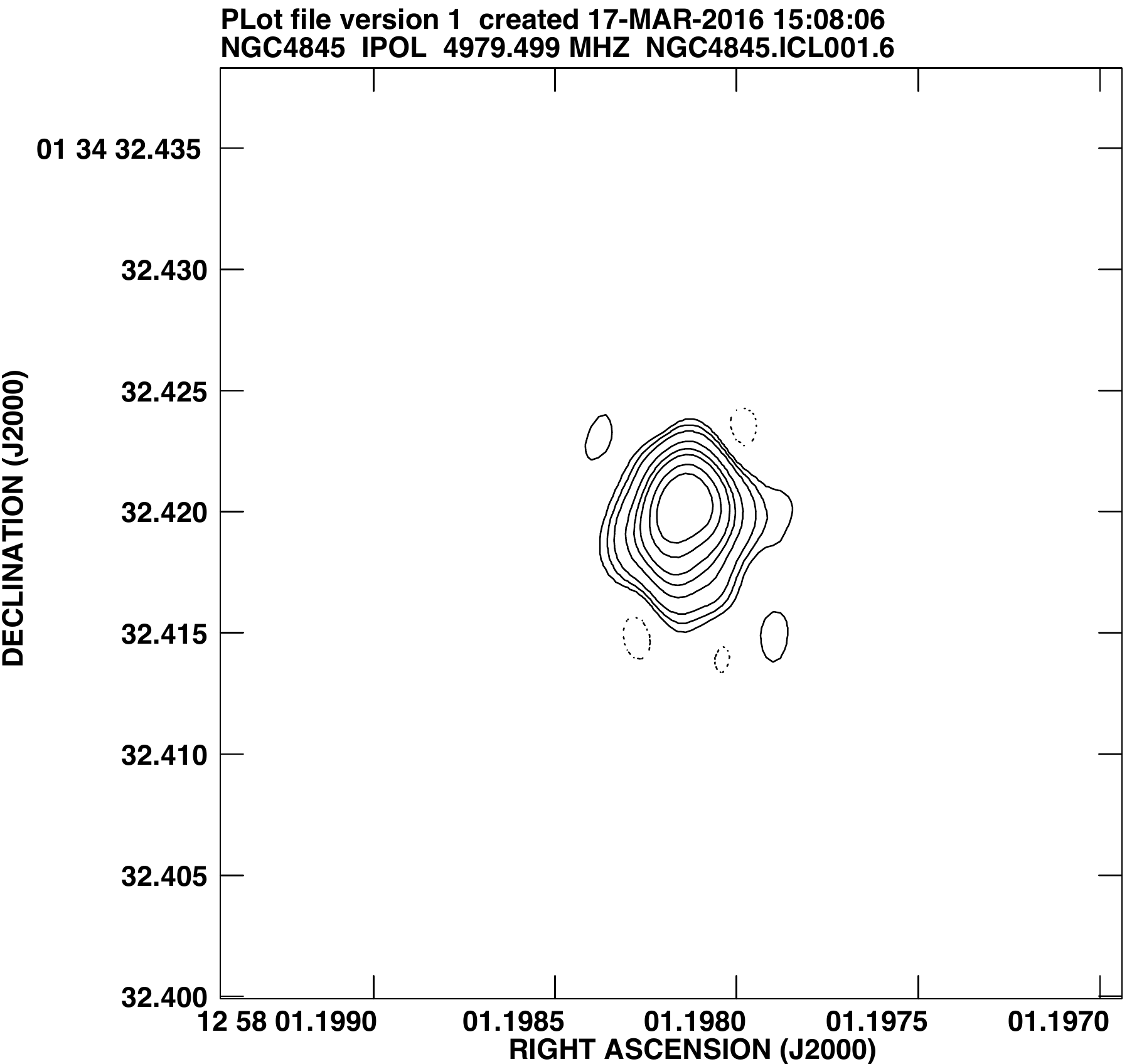}

~~~~~~~~~\includegraphics[width=7.6cm]{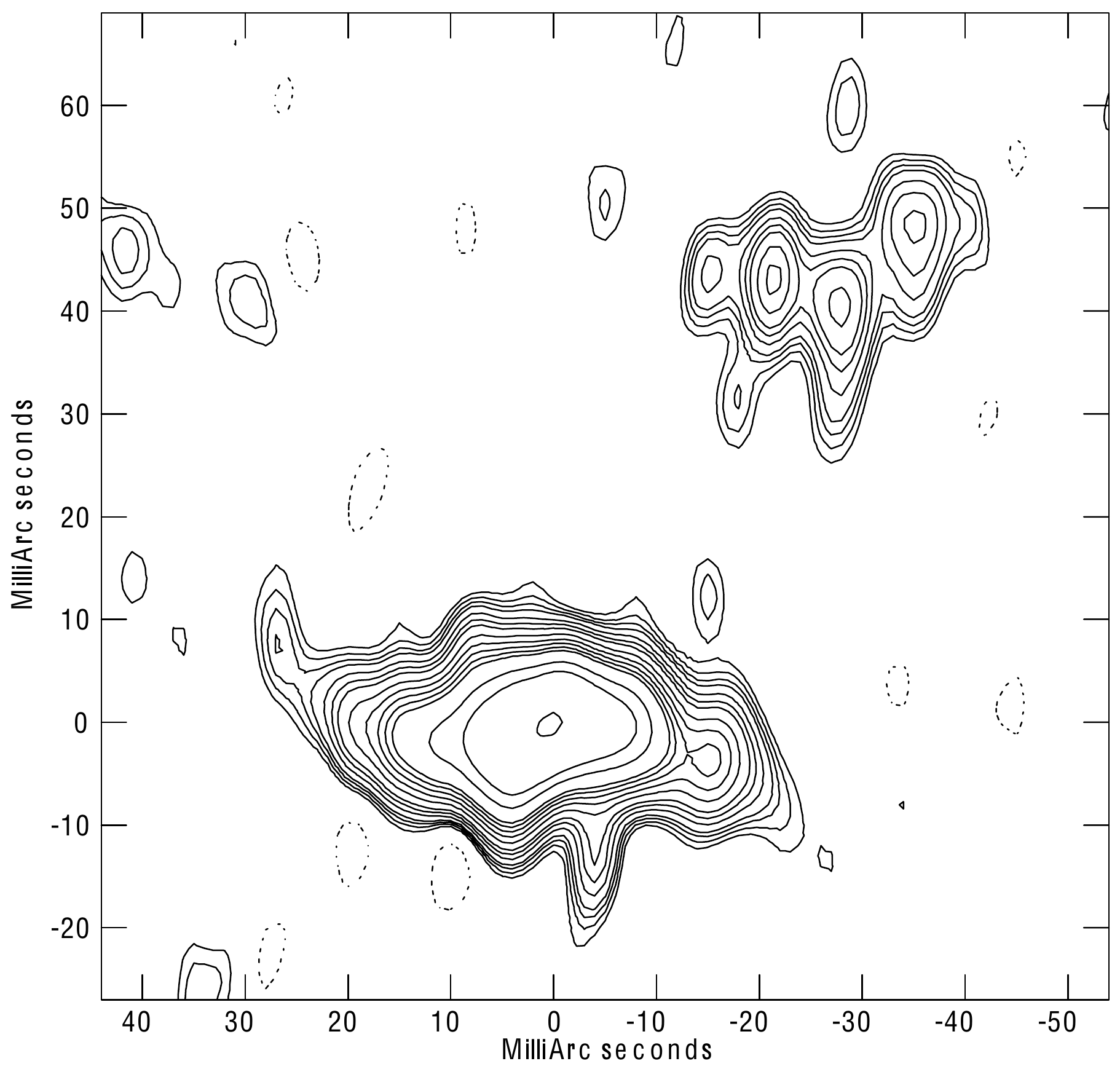}
\caption{The nuclear region of NGC 4845, as seen in 2015 with the {\sl VLBA} at (top) C band and (bottom) L band.
The contours in both panels are at (-3, 3, 5, 8, 16, 24, 32, 48, 64, 128, 256) times a base contour level of 
0.1 mJy/beam.  The RMS noise of the image is given in Table 1. In L band, we have assumed that the position of the 
flux maximum is at (0,0)}
\end{figure}
The  NW source is seen only at L band.  
We do not use the RMS noise of the {\sl VLBA} data to put an upper limit
on its C-band flux, as that source is located more than 20 
beam-widths away from the nuclear source. The task of detecting the northern source at 
C band is made even more difficult by its extended nature (Figure 3), spread over several beam-widths in the 
L band data, with a significance peaking at just over $8 \sigma$ in those data.  The combination of those factors
means that if the NW source had $\alpha \approx 1$, a fairly steep spectrum typical of the mini-lobes seen 
in GPS sources (e.g., Snellen et al. 1999, Tzioumis et al. 2002), 
it would be difficult or impossible to detect on our C band {\sl VLBA} image.

\subsection{Polarized Emission}

A very interesting property of the radio source is its circular polarization, first noted in the early CHANG-ES 
observations by Irwin et al. (2015).  As shown in Table 2, our {\sl VLBA} observations reveal a significant ($\sim 10 
\sigma$) detection of circular 
polarization in L band, while no circular polarization is seen in the {\sl JVLA} data or at 
C band in the {\sl VLBA} observations.  We do not detect any significant linear polarization in any of our images
(Table 2).   We show an image of the circular polarization (Stokes $V$) in Figure 5.  As seen, the circularly 
polarized emission in L band is resolved, but given the fact that it lies southeast of the higher frequency, C band flux 
maximum (see Figures 4 and 5), which in AGN is usually interpreted as more indicative of the location of the nucleus, 
its apparent alignment with the position angle between the central and NW source is 
likely an illusion.

We used JMFIT to measure the flux 
and extent of this component 
The resolved component has a
size of $11.0 \pm 0.7 \times 4.5 \pm 0.3 $ milliarcsec in PA 162$^\circ$.  However, due to the low flux of 
the component, JMFIT was not able to converge well on a deconvolved size for this component, 
obtaining $6.7 \pm 1.6 \times 0.0 \pm 0.5$ milliarcsec in PA 138$^\circ$.   
As can be seen, the level of circular polarization in our 2015 {\sl VLBA} data 
is somewhat decreased from the $\sim 3\%$ values 
seen in L band epochs  $T_1 - T_3$.
We discuss further the evolution and nature of the polarization in \S 4.4.

\begin{figure}[ht]
\centerline{\includegraphics[width=7.6cm]{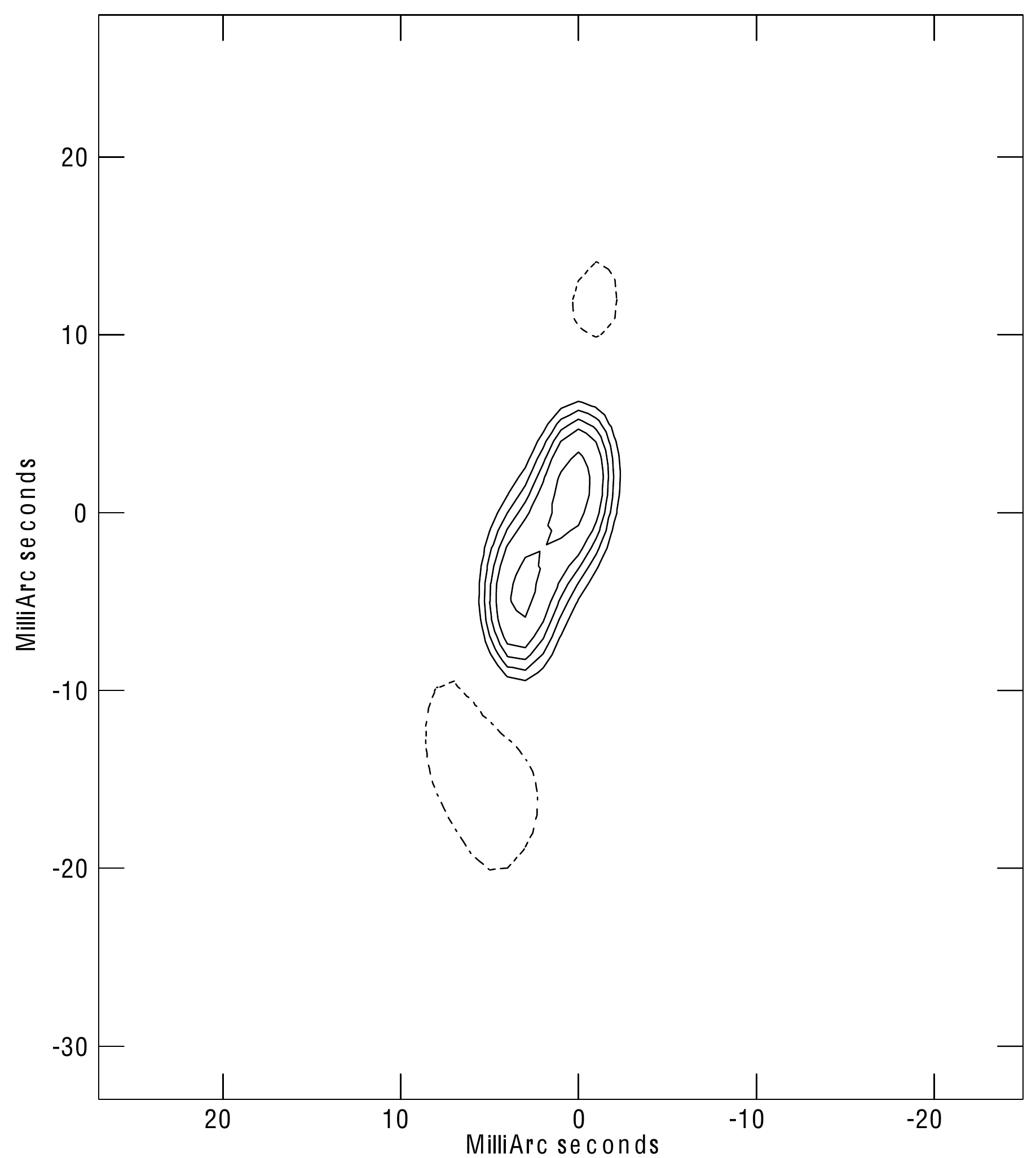}}

\caption{The nuclear region of NGC 4845, as seen in Stokes $V$ in 2015 with the {\sl VLBA} at L band.  The contour
levels are identical to Figure 4, and we have assumed the flux maximum is at (0,0).}
\end{figure}

\section{Discussion}
 
The disruption of a star or substellar object by tidal forces in the neighborhood of a galaxy's SMBH is an exciting
event that has broad-ranging implications.  Large amounts of material can be injected into the accretion flow 
surrounding the black hole, and if the previous rate of accretion was small, an inactive black hole 
can become active -- and in the process, exhibit some of the same properties as viewed in other, less transient 
AGN classes.  However, not all AGN properties may be exhibited in a TDE.  The properties seen 
may depend on the exact physical conditions during the TDE as well as the nuclear environment.
However there are few unambiguous detections in the radio of a compact source connected with a TDE.  
Zauderer et al. (2011) reported {\sl JVLA} observations of the TDE Swift J1644+57.
That source, further analyzed by Berger et al. (2012) and Zauderer et al. (2013), appeared to occur in a 
fairly pristine galactic environment, and appeared to decline precipitously in brightness after several 
hundred days.  The latter ``core shutoff" (Zauderer et al. 2013) has also likely occurred for IGR12580+0134
(Nikolajuk et al. 2013).  Swift J1644+57 was observed also by the EVN 
(Yang et al. 2016), which found a compact radio source but no evidence for superluminal motion.

Romero-Ca\~nizales et al. (2016) claim the detection of a compact milliarcsec nuclear source,
 in the TDE ASASSN-14li.   It was hosted in the post-starburst 
galaxy PGC 04324 (d=90 Mpc), and the radio source they identify is far fainter ($\sim 1$ mJy at L band as 
compared to over 60 mJy in our observations).  Thus the nuclear source we detect in IGR12580+0134
is over a factor 2 more 
powerful and our constraints on its size (using the C band FWHM)  are a factor 
6 smaller -- corresponding to a  projected linear size 
of $<0.16$ pc, although we caution the reader that the nuclear source is resolved.  Thus, while future, higher 
frequency observations can constrain this size further, we can say that   the observed radio
structure is not consistent with non-relativistic, 
spherical expansion of the disrupted object's remains.  An alternative scenario that might 
produce a roughly spherical source is the scattering of the high-energy radiation from the TDE by nuclear dust.
Lu et al. (2016) showed that this could produce significant near-infrared and mid-infrared emission for months to
years after the event.  While a possibly exciting prospect for JWST,  
such emission would be undetectable in the radio. Also unlikely, for similar reasons (plus the fact that a bright 
flare of a factor of ten would be difficult to explain) is a bright source connected with the Pa $\alpha$ emission
from the nuclear mini-spiral, discussed by Wang et al. (2010).
In this section, we discuss the implications of our findings, both in terms 
of the nature of the milli-arcsecond resolved component and its evolution, as well as the evolution of the broadband radio 
spectrum and the source polarization. 

\subsection{Nature and Evolution of the Milli-arcsecond Scale Emission}

The discovery of a resolved component in {\sl VLBA} imaging of IGRJ12580+0134 is exciting, and deserves a full 
discussion of the implications. There are two possibilities.  The first is that the NW component is unassociated 
with the TDE.  We consider this possibility (which we consider more likely) at the end of this section.  However, an 
alternate, more exciting possibility needs to be considered.  If one assumes that the northwestern source 
constitutes material ejected during the TDE in a relativistic flow, the average implied speed of
the component is  $\beta_{app} = 2.4c$.   This value of $\beta_{app}$  is not inconsistent with the estimate of viewing angle $\theta=40^\circ$ from Lei et al. (2016), which would
indicate an ejection velocity of 0.96$c$ and a bulk Lorentz factor $\Gamma=3.6$.  At one level, this scenario seems likely, and the morphology of the NW {\sl VLBA} source is suggestive of an 
expanding source at the working surface of a conical outflow.
The implied bulk Lorentz factor is 
significantly lower than the $\Gamma=10$ assumed in the model of Lei et al. (2016). 
This is not problematic.  The combination of $\beta$ and $\theta$  is
not robustly constrained, and in addition, if the jet is interacting with a sufficiently dense
CNM, it is very likely that its working surface would decelerate as a function of time (see \S 4.2)

However, we believe it is most likely that 
the NW component is not connected with the 2010 November TDE, because (as we detail in \S 4.2) we 
consider it likely that the interaction of the jet with the dense CNM produced significant deceleration in the jet within
the first year. Such an interpretation (which will be tested by second-epoch {\sl VLBA} observations in Spring 2017) is also suggested by 
the orientation of the extension to the circularly polarized L band source, which (taking the point closest to the C 
band nucleus as "upstream") points {\it away} from the NW source.  This is not the first resolved component seen in 
a TDE.  Romero-Ca\~nizales et al. (2016) also claimed the detection of a 
fainter (0.1 mJy), possibly elongated companion source 4.3 milliarcsec (1.9 pc projected distance) from the 
nuclear source of ASSASN-14li. That component, like the NW source here, could either be a jet component 
related to the TDE, the remnant of a previous TDE event, or a binary black hole.  

If the NW component is unrelated to the 2010 TDE, it could 
either be a remnant of an earlier TDE event or  a supernova remnant
of luminosity $\sim 10^{27} {\rm erg~ s^{-1} ~Hz^{-1}}$ (given the resolved nature of the NW component, a binary black hole origin is extremely unlikely). While uncommon, that is not  exceptional for supernova 
remnants in nearby galaxies (see, e.g., the discussion and luminosity functions in Chomiuk \& Wilcots 2009).  Moreover,
this latter interpretation is consistent with the fact that NGC 4845, the host of this TDE, is a dusty, spiral galaxy that 
hosts a LINER\citep{spin89}, and perhaps also active star formation in its center.  It could also be a smaller-scale 
counterpart to the large-scale disk seen on the {\sl JVLA} images, although this seems less likely, as 
such a high surface brightness feature (total extended flux in L band is greater than the unresolved
component) would be difficult to reconcile with the low activity state of the
nuclear black hole prior to the TDE. More likely is a SNR origin,  although the FIRST and
NVSS fluxes set an upper limit on the flux from SNRs.

\subsection{A Model for the Expanding Nuclear Structure}

On a more detailed level,
the physical picture of the jet evolution in the CNM is similar to that of GRBs, except for an off-axis viewing angle
correction \citep{2002ApJ...570L..61G}. The central engine, the tidal 
disruption of stellar objects by SMBHs, powers relativistic jets, which 
then propagate in the CNM. Shocks can be produced due the the jet-CNM 
collision, and high energy electrons can be accelerated. The synchrotron 
and/or inverse Compton emission of accelerated electrons gives the 
multi-wavelength afterglow emission. 

Here we model the possible evolution of the ejecta. The evolution of the jet 
can be roughly divided into three stages, the coasting phase, the 
deceleration phase, and the Newtonian phase. Its dynamics is governed by 
a set of hydrodynamical equations \citep{2000ApJ...543...90H}. We solve the
dynamics of the forward shock numerically, and calculate the synchrotron
emission from the accelerated electrons \citep{1998ApJ...497L..17S}.
The parameters of the model include the launching time, energetics, 
initial Lorentz factor, opening angle, electron and magnetic energy 
partition of the jet, the spectral index of accelerated electrons, 
the viewing angle, and the CNM density (see Table \ref{table:para}).
We assume an instantaneous
injection of energy into the jet.

\begin{figure}[!htb]
\centering
\includegraphics[width=1.0\columnwidth]{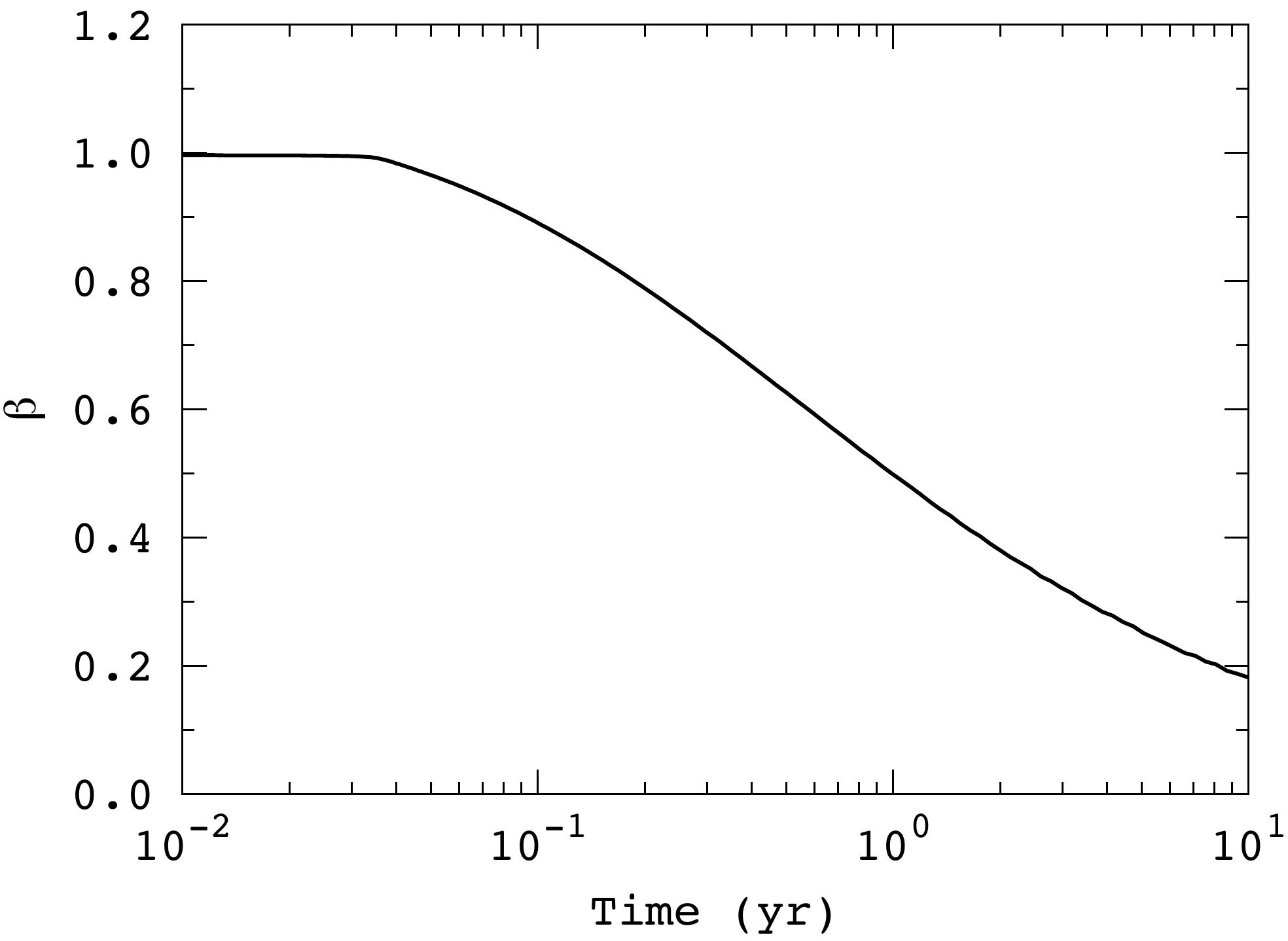}
\caption{Evolution of the velocity of the jet with time in the observer's
frame.}
\label{fig:beta}
\end{figure}

\begin{deluxetable*}{ccccccccc}
\tabletypesize{\scriptsize}
\tablecolumns{8}

\tablewidth{0pt}
\tablecaption{Model Parameters (see \S 4.2).}
\tablehead{
\colhead{Jet Launching Time} & \colhead{CNM Density} & \colhead{Viewing Angle} & 
\colhead{Jet KE} &\colhead{Initial} & \colhead{Initial} \\
\colhead{$\Delta t$ (day)} &  \colhead{$n ({\rm cm}^{-3})$} & \colhead{$\theta_{\rm obs}$ (deg)} & 
\colhead{$E$ ($\times 10^{50}$ erg s$^{-1}$)} &  \colhead{$\Gamma_j$} & \colhead{$\theta_j$ (deg)} & 
\colhead{$p^a$} & \colhead{$\epsilon_e^b$} & \colhead{$\epsilon_B^b$} }
\startdata
 18$^c$ & 1.2 & 35 & 530   & 11.2 & 3.7 & 2.70 & 0.21 & 0.05 \\
\enddata
\tablenotetext{a}{Spectral index of accelerated electrons}
\tablenotetext{b}{Fraction of the ejecta kinetic energy assigning to accelerated electrons or the magnetic field}
\tablenotetext{c}{Relative to March 25.5, 2011}
\end{deluxetable*}
\label{table:para}

The deceleration time of the jet can be estimated as
\begin{eqnarray}
t_{\rm dec}&=&a_{\rm off}^{-1}(1+z)\left[\frac{3E_{k}}{16\pi nm_p\Gamma^8c^5}
\right]^{1/3} \nonumber\\
&\simeq&0.13\,a_{\rm off}^{-1}n^{-1/3}E_{50}^{1/3}\Gamma_1^{-8/3} \,{\rm days},
\end{eqnarray}
where $E_{50}=E_k/10^{50}\,{\rm erg}$, $\Gamma_1=\Gamma/10$. The off-axis
factor $a_{\rm off}$ is defined as the ratio of the off- to on-axis 
Doppler factor, $a_{\rm off}=(1-\beta)/(1-\beta\cos\psi)$, with $\beta$ 
being the velocity of the jet in units of the light speed, and $\psi=
\max(\theta_{\rm obs}-\theta_j,0)$ being the angle between the jet moving
direction and the line-of-sight. 
For the parameters given in Table 3, $t_{\rm dec}\sim10$ day. The integral distance traveled 
by the jet is about $0.4$ pc, which is much smaller than the separation 
between the NW and SE sources ($\sim4.1$ pc).  This is roughly consistent with the size 
of the circularly polarized component (Fig. 5), 
The average implied speed of the ejecta over five years would then be $\sim 0.3c$.

Other TDEs have also had models published for subluminally modeled components.  
The upper limit published by Yang et al. (2016)  for the 
average velocity of Sw 1644+57's ejecta was similar to ours.  
%
%
Alexander  et al. (2016) modeled a subluminally moving component in ASSASN-14li that could be traced back to 
the TDE itself, using the radio spectrum alone.  The possibility of significant 
unbound matter (perhaps as much as half of the original mass of the disrupted object) was discussed by 
Krolik et al. (2016), who pointed out that if left undisturbed
the material would coast outward from the black hole at a speed $\sim [GM_{\rm BH}/a_{\rm min}]^{1/2}$, where
$a_{\rm min}$ is the semi-major axis of the disrupted object's original orbit. 
And, interestingly, Giannios \& Metzger (2011) and Mimica et al. (2015) have published a model for Sw 1644+57 
that includes an ultrarelativistic core (Lorentz factor $\Gamma \sim 10$) surrounded by a slower ($\Gamma \sim 2$) 
sheath that provides a reasonable fit to that TDE's lightcurve.  That model also would fit the lightcurve of IGR 12580+0134, but in NGC 4845's nuclear environment it would likely be subject to the same deceleration that we discuss above.

\begin{figure}[!htb]
\centering
\includegraphics[width=1.0\columnwidth]{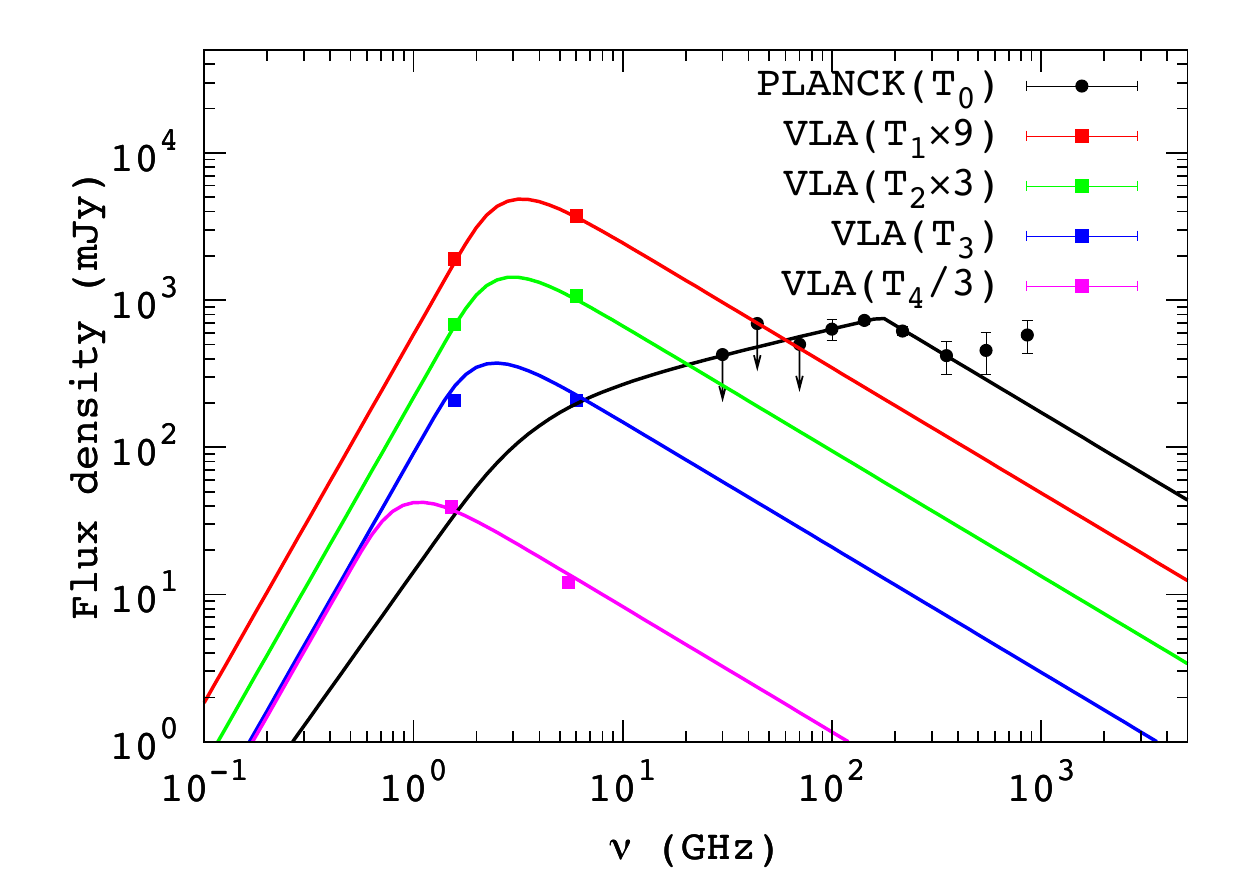}
\caption{
Spectral energy distribution of the jet emission  from {\sl JVLA} observations at L-band ($\sim1.54$ GHz) and C-band ($\sim5.75$ GHz), and 
observed by {\sl Planck} at $T_0=T_1-348$ days \citep{2016MNRAS.461.3375Y}, 
and by {\sl VLA} at $T_1$, $T_2$, $T_3$, and $T_4$. For visual clarity, 
the SEDs have been vertically offset, as indicated in the legend.  Lines show the model fitting
results.  See \S 4.2 for details.
}
\label{fig:fitting}
\end{figure}

\subsection{Evolution of the Circumstellar Medium and Broadband Spectrum}

Lei et al. (2016) and Irwin et al. (2015) fitted the early {\sl VLA} data with a 
jet-CNM interaction model that assumes a conically expanding, 
synchrotron emitting jet source that is initially optically
thick due to synchrotron self-absorption at gigahertz frequencies, with an optical thickness that changes as a 
function of time, likely as a result of the expansion of the jet.  The model assumed an initial Lorentz factor
of the jet as $\Gamma_i\sim10$ and a viewing angle of $40^{\circ}$.  
Fig. \ref{fig:fitting} shows the SEDs for 
a joint fitting to the {\sl VLA} observations at four epochs, $T_1$ (2011-12-30),
$T_2=T_1+56$ days, $T_3=T_1+196$ days, and $T_4=T_1+1270$ days
\citep{irwin2015}, as well as the PLANCK observations at 
$T_0=T_1-348$ days \citep{2016MNRAS.461.3375Y}. Model parameters
are given in Table 3. 

The flux densities at different
epochs can be well described by the model.  The decline in the source flux follows the $t^{-5/3}$ behavior predicted by Irwin et al. (2015, their Eqn. 22).
Earlier observations showed that the spectrum peaked between L-band and C-band with the peak shifting to 
lower frequencies with time (Irwin et al. 2015).  We now see a continuation of that spectral trend, i.e. that the peak 
of the spectrum has shifted to a frequency below L-band.  Note also that the shape of the radio 
through IR 
SED rules out a significant thermal contribution except at epoch $T_0$ when the submm upturn in the Planck data
could be explained by thermal dust emission from a torus, perhaps heated by the TDE.  This would also be 
 reasonable given its classification as a LINER \citep{spin89}.

\begin{figure}[!ht]
\centering
\includegraphics[width=1.0\columnwidth]{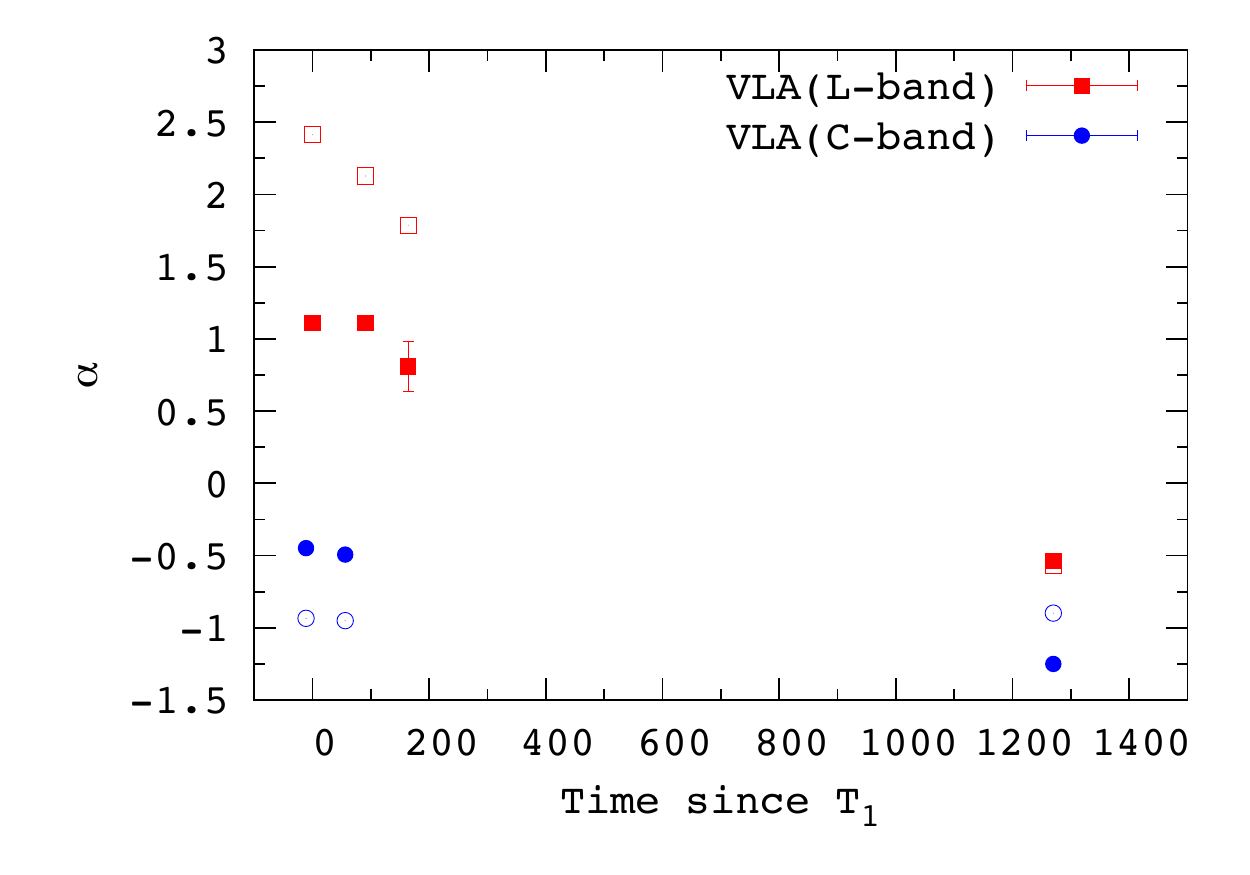}
\caption{In-band spectral indices of the TDE, from the {\sl VLA} measurements
(filled symbols) at L- and C-bands, compared with that expected from the 
jet-CNM model (open symbols).
}
\label{fig:index}
\end{figure}  

Fig. \ref{fig:index} shows the comparison of the in-band spectral indices
between the measurements and the model prediction. We find that the model 
prediction can roughly reproduce the evolution trend (i.e., from hard to soft) of the spectra. 
However, quantitatively there are discrepancies between these two.   Some of this is due to the 
details of the model (shock strength, injection index and Mach number, for example).  In addition,
at L-band, the emission is significantly affected by the self-absorption 
of the synchrotron emission, which makes the estimate of the spectral 
indices inaccurate. At C-band, however, the emission is expected to be 
optically thin, and the evolution of $\alpha$ may indicate spectral 
variations of the accelerated electrons during the jet propagation.

  \subsection{Interpretation of the Circular Polarization}
  
The lack of detection of circular polarization at C band in our {\sl VLBA} data
is consistent with the earlier {\sl JVLA} observations from Irwin et al. (2015).  Those authors explained the circular 
polarization observed as due to conversion from linear polarization to circular polarization via generalized 
Faraday rotation (see also O'Sullivan et al. 2013, Beckert \& Falcke 2002).  
The decrease of the circular polarization at L band over five years
could then be explained via a gradually decreasing Faraday depth, consistent with the declining optical 
depth that explains the gradual spectral evolution shown in Figure 3.   A closer look at Figure 3 shows that not only has 
the source spectrum has taken on a much more power-law type shape over
the intervening five years since epoch $T_1$, but in fact, as discussed in \S 3.1, this indicates a low optical depth 
for the source at epochs $T_4$ and later on large scales 
including the time of our {\sl JVLA} observation. 
In that case, we would not expect to see a significant circular 
polarization. However, on small scales it might still be possible to see small optically thick portions in the 2015 {\sl VLBA}
observations.  We regard this as more likely than the alternate explanation of variability for the lack of arcsecond-scale 
circular polarization in the 2015 {\sl JVLA} observations.  While plausible, as Hovatta et al. (2012) and Homan et al. 
(2006) have pointed out that circular and linear polarization vary commonly for MOJAVE blazars, it does require 
additional complexity and therefore Occam's razor argues against it.
Perhaps most critically, though, we note that the C band flux maximum lies at the northern end of the 
circularly polarized region in L band. If indeed the C band flux comes predominantly from downstream of the L band 
flux maximum then  the C-band emitting region has an inverted 
and optically thin spectrum more similar to many VLBI cores.  The future development of this region will be 
interesting to watch.

\subsection{Conclusions}

The TDE IGRJ 1258+0134 is now seen to have  a complex radio structure on milliarcsecond scales that is most likely 
connected to the TDE. 
The radio spectrum displays a complex evolution that, while broadly consistent with earlier modeling 
of this source, is much better defined with the addition of these data.
The nature of the parsec-scale structure is both unclear.  While the observed 
NW component could be evidence of a superluminally moving component ejected as part of the CMB, 
we believe this 
is unlikely, as modeling suggests that due to the dense nuclear ICM the jet should decelerate on timescales of 
months and have a much smaller, sub-parsec size, more consistent with the size of the circularly polarized component
observed at L band.

\begin{acknowledgments} 
Based on observations made with the Karl G. Jansky Very Large Array ({\sl JVLA}) and the Very Long Baseline array ({\sl VLBA}).  
We acknowledge an interesting conversation with Sjoert Van Velzen about this paper.
\end{acknowledgments}

\end{document}